\documentclass{iau}
\usepackage{graphicx}

\title[Looking for WHIM] 
{Finding and characterising WHIM structures \\
using the luminosity density method}

\author[J. Nevalainen et al.]   
{Jukka Nevalainen$^1$
\and L. J. Liivam\"agi$^1$
\and E. Tempel$^1$
\and E. Branchini$^2$
\and M. Roncarelli$^3$
\and C. Giocoli$^3$
\and P. Hein\"am\"aki$^4$
\and E. Saar$^1$
\and M. Bonamente$^5$
\and M. Einasto$^1$
\and A. Finoguenov$^5$
\and J. Kaastra$^6$
\and E. Lindfors$^4$
\and P. Nurmi$^4$
\and Y. Ueda$^7$}

\affiliation{$^1$Tartu Observatory, Observatooriumi 1, 61602 T\~oravere, Estonia \\ 
email: {\tt jukka@to.ee} \\[\affilskip]
$^2$University Roma Tre, via della Vasca Navale 84, 00146 Roma, Italy \\ [\affilskip]
$^3$University of Bologna,  viale Berti Pichat 6/2, I-40127 Bologna, Italy \\  [\affilskip]
$^4$Tuorla Observatory, V\"ais\"al\"antie 20, FI-21500  Piikki\"o, Finland \\ [\affilskip]
$^5$University of Alabama in Huntsville, Huntsville, AL 35899, USA \\[\affilskip]
$^6$SRON Netherlands Institute for Space Research, Sorbonnelaan 2, 3584 CA Utrecht, the Netherlands \\[\affilskip]
$^7$Kyoto Observatory, Yoshida-honmachi, Sakyo-ku, Kyoto 606-8501 JAPAN
}

\pubyear{2014}
\volume{308}  
\pagerange{119--126}
\jname{The Zeldovich Universe}
\editors{ R. van de Weygaert, S. Shandarin, E. Saar \& and J. Einasto, eds.}
\begin{document}

\maketitle

\begin{abstract}
We have developed a new method to approach the missing baryons problem. We assume that the missing baryons reside in a form of Warm Hot Intergalactic Medium, i.e. the WHIM. 
Our method consists of (a) detecting the coherent large scale structure in the spatial distribution of galaxies that traces the Cosmic Web and that in hydrodynamical simulations is associated to the  WHIM, (b) map its luminosity
into a galaxy luminosity density field, (c) use numerical simulations to relate the luminosity density to the density of the WHIM, (d) apply this relation to real data to trace the WHIM using the observed galaxy luminosities in the Sloan Digital 
Sky Survey and 2dF redshift surveys. In our application we find evidence for the WHIM along the line of sight to  the Sculptor Wall, at redshifts consistent with the recently reported X-ray absorption line detections. 
Our indirect WHIM detection technique complements the standard method based on the detection of characteristic X-ray absorption lines, showing that the galaxy luminosity density is a reliable signpost for the WHIM. For this reason, 
our method could be applied to current galaxy surveys to optimise the observational strategies for detecting and studying the WHIM and its properties. Our estimates of the WHIM hydrogen column density N$_H$ in Sculptor agree with those obtained via the X-ray 
analysis. Due to the additional N$_H$ estimate, our method has potential for improving the constrains of the physical parameters of the WHIM as derived with X-ray absorption,
and thus for improving the understanding of the missing baryons problem.

\end{abstract}

\section{Introduction}
At low redshifts (z$<$2) all observations of the visible matter sum up only to $\sim$70\% of the expected cosmological mass density of baryons (e.g. \cite[Shull et al., 2012]{Shull12}). Large scale structure formation simulations suggest that these missing baryons reside in the form of Warm-Hot Intergalactic Matter (WHIM)  in the filamentary 
structure connecting the clusters of galaxies and superclusters (e.g. Cen \& Ostriker, 1999, 2006). At the predicted temperatures of $10^5 - 10^7$ K and densities $10^{-6} - 10^{-4}$ cm$^{-3}$ the X-ray emission from single WHIM structures is too faint to be detected with current instrumentation.
However, the column densities of the highly ionised WHIM metals along  large--scale filamentary structures can reach  a level of $10^{15} - 10^{16}$ cm$^{-2}$, imprinting detectable absorption features on the soft X-ray spectra of background sources (e.g. Nicastro et al., 2010).

While the X-ray absorption measurements are crucial in the analysis of WHIM, they are sparse. The usual "blind search" is based on observing random bright blazar flares, without a priori knowledge of the foreground structure.
To improve this, we focus on \emph{known foreground large scale structures} traced by galaxies, 
like the cosmic filaments, that most likely contain WHIM. We have applied the Bisous model (\cite[Stoica et al., 2005]{Stoica05}) to trace and extract the filamentary network in the presently largest galaxy redshift
survey, SDSS DR8 (\cite[Tempel et al., 2014]{Tempel14}) and 2dF. Thus we have an extensive data base of potential WHIM structure locations, sizes and redshifts. 

We report distances and lengths in co-moving coordinates (unless stated otherwise), using $\Omega_m$ = 0.3,  $\Omega_{\Lambda}$ = 0.7  and H = 70 km s$^{-1}$ Mpc$^{-1}$

\section{The method} 
\subsection{Luminosity density fields} 
We have examined the above galaxy structures using the luminosity density (LD) method, pioneered by our group (e.g. Liivam\"agi et al., 2012). With this method we quantify the 3-dimensional galaxy distribution using the optical (R - band) light in galaxies. We assume that every galaxy is a visible member of a density enhancement (group or cluster). We place the galaxies at the mean distance of the group or the cluster, to correct for the effect of the dynamical velocities (finger-of-god). 
We correct the galaxy luminosities for the observational magnitude limited samples by a weighting factor that accounts for the group galaxies outside the visibility window. The galaxy luminosity distribution is then smoothed with a  B3-spline kernel function. The smoothing length determines the characteristic scale of the objects under study. We have adopted 1.4 Mpc as the smoothing scale, in order to match the filament widths. We then sample the LD distribution at the points of a uniform grid, encompassing the survey volume,  with a sampling scale of 1.4 Mpc, thus creating the LD field. For a given galaxy structure, we use the LD field to evaluate the LD profile (see Fig. 1).

\begin{figure}[h]
\vspace*{-0.5cm}
\begin{center}
\includegraphics[width=3.0in]{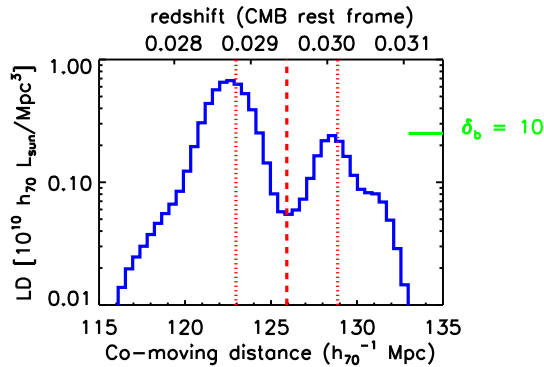} 
\caption{Luminosity density (LD) profile along the Sculptor Wall galaxy structure at z$\sim$0.03 as obtained from the 2dF data is shown with a (blue) solid line.
The best-fit value and the 1$\sigma$ uncertainties of the redshift of the X-ray absorber (Fang et al., 2010) are shown in (red) dashed and dotted lines.
Note that the redshifts and distances are reported in CMB rest frame. The luminosity density value corresponding to baryon overdensity  $\delta_b$ = 10, as estimated with Eq. \ref{LD-WHIM.eq}, is denoted with (green) symbol on the right axis.}
\label{fig1}
\end{center}
\end{figure}

\subsection{Large Scale Structure simulations} 
If the galaxies follow the underlying dark matter (DM) potential, similarly as  the WHIM, then the luminosity density field in filaments can be used to trace the missing baryons.
To test the validity of this assumption we have applied our filament--finding and LD field algorithms to a distribution of mock galaxies, DM  and the WHIM (i.e. gas at temperatures 10$^5$ - 10$^7$ K) at z = 0 in the hydrodynamical simulations (\cite[Cui et al., 2012]{Cui12}).
These simulations make use of smoothed particle hydrodynamics in GADGET-3 code to produce dark matter and diffuse baryonic components within a box with a size of 570  Mpc.
The simulations involve radiative cooling, star formation and feedback from supernova remnants.  The galaxies are created by populating the DM  haloes using
a halo occupation distribution constrained with the SDSS data (\cite[Zehavi et al., 2011]{Zehavi11}). The simulated data are adequate to follow the different density components with a resolution of $\sim$1 Mpc.
  
Indeed the LD and WHIM density ($\rho_{WHIM}$) in the above simulations correlate rather well (Pearson correlation coefficient = 0.85). We used this correlation to derive a quantitative relation between these, i.e. the LD - $\rho_{\rm WHIM}$ relation,
within the filaments identified by the mock galaxies (see Eq. \ref{LD-WHIM.eq}  and Fig. 2). 
\begin{eqnarray}
\rho_{WHIM} \approx  63 \times LD^{1.6}, when \ LD < 0.35  \nonumber \\
\rho_{WHIM} \approx  33 \times LD^{1.0}, when \ LD > 0.35,  
\label{LD-WHIM.eq}
\end{eqnarray}
where LD and $\rho_{WHIM}$ are expressed in units $10^{10}  \ h_ {70}  \  L_{\odot}  \ Mpc^ {-3}$ and $10^{10} \ h_ {70}^ {2} \ M_{\odot} \ Mpc^ {-3}$.

\begin{figure}[h]
\vspace*{-1.0 cm}
\begin{center}
\includegraphics[width=4.8in]{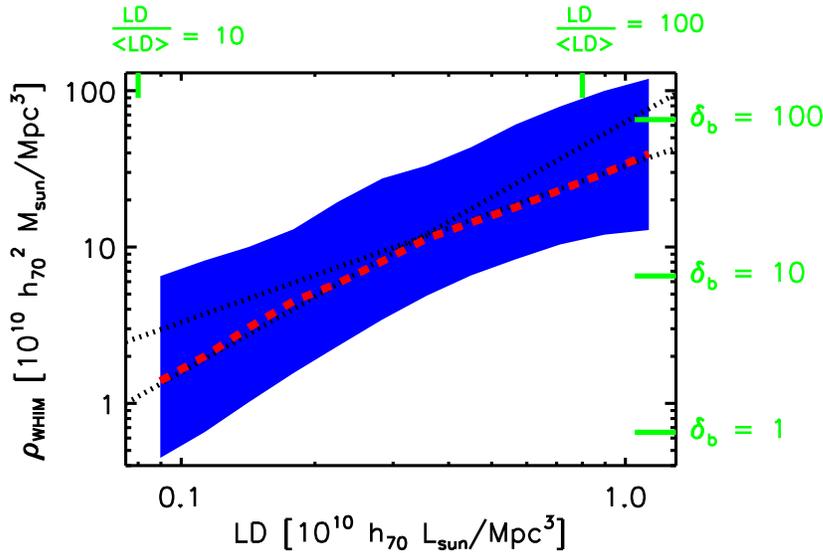} 
\vspace*{-0.5 cm}
\caption{The luminosity density (LD) - WHIM density ($\rho_{WHIM}$) relation, and the 1$\sigma$ uncertainties derived from the cosmological hydrodynamical simulations (Cui et al., 2012) are indicated with (red) dashed line and the (blue) shaded region, respectively.
The black dotted lines indicate the power-law approximations to the relation (Eq. \ref{LD-WHIM.eq}). The (green) symbols indicate different levels of overdensity of baryons and luminosity.}
\label{fig2}
\end{center}
\end{figure}
We use this relation to convert the observational (SDSS and 2dF) LD values into $\rho_{WHIM}$ estimates. We then integrate the $\rho_{WHIM}$ profile of a given filament to obtain the 
WHIM hydrogen column density $N_{\rm H}$. The observed LD profile values typically correspond to baryon overdensities of $\sim$ 10 (see Fig. 1), consistent with those predicted by simulations for the WHIM filaments (e.g. Cen \& Ostriker 1999, 2006). 

\section{Results}
In order to test the feasibility of our WHIM $N_{\rm H}$ estimation method, we extracted the LD profile along the line-of-sight to the blazar H2356-309 behind the Sculptor Wall, where X-ray measurements of WHIM absorption have been obtained  (Fang et al., 2010; Buote et al., 2009, Zappacosta et al., 2010). We found $\sim$10 Mpc long luminosity density structures at redshifts  consistent with those measured  with X-rays (see Fig. 1).
Our LD-WHIM density relation yields WHIM $N_{\rm H}$ values consistent with those measured in X-rays (see Table \ref{NH.tab}), proving the reliability of our column density estimation method.

\begin{table}
\begin{center}
\caption{Redshifts and WHIM hydrogen column densities $N_{\rm H}$ in Sculptor}
   \smallskip
   \label{NH.tab}
{\scriptsize
\begin{tabular}{lcccc}\hline \vspace{2mm}
                                               &    X-ray$^1$      & LD$^2$                 &      X-ray$^1$            & LD$^2$                                            \\
system                                    &  redshift$^3$    &  redshift$^3$          &  $\log{N_{\rm H}}$    & $\log{N_{\rm H}}$ \\  \hline 
Sculptor Wall (SW)                  &  0.030--0.032   &   0.028--0.033       & 21.0[20.1--22.3]        & 20.0[19.6--20.6]     \\
Pisces-Cetus (PC)                   &  0.060--0.063   &   0.060--0.063       & 20.1[19.9--20.3]        & 19.8[19.4--20.3]     \\
Farther Sculptor Wall (FSW)    &  0.125--0.127   &   0.128--0.129       & 20.8[20.0--21.2]        & 19.9[19.6--20.5]     \\ \hline
\end{tabular}
}
\end{center}\vspace{1mm}
\scriptsize{
{\it Notes:}\\
$^1$Redshift centroid (Chandra) and WHIM $N_{\rm H}$ estimates based on X-ray absorption measurements (SW: Fang et al., 2010 ; PC and FSW: Zappacosta et al., 2010). The $\log{N_{\rm H}}$ values are derived assuming O abundance 0.1 Solar, and O/H ratio from Grevesse \& Sauval, 1988.
The estimate for SW is obtained here by assuming T = $10^6$ K.   \\
$^2$Our luminosity density - based estimates for the redshift range corresponding to the extent of the given system intersected by the H2356-309 sightline, and WHIM $N_{\rm H}$ of the given system.}\\
$^3$In the observational frame.
\end{table}

\section{Discussion}
Our plan is to apply our method to current galaxy surveys (and e.g. to Euclid results in near future) to optimise the observational strategies for detecting and studying the WHIM and its properties.
In particular, we will cross-correlate the significant WHIM structures found from e.g. SDSS and 2dF with bright background blazars.
We aim at obtaining XMM-Newton/RGS and Chandra/LETGS data of such blazars in a high flux state, which are located in the line-of-sight to WHIM structures with the highest WHIM column density estimates. 
This will be useful also for next generation X-ray telescopes like ATHENA. We will also extend this work to include far-ultraviolet WHIM measurements with e.g. FUSE and HTS of the low temperature WHIM.

{}

\end{document}